\documentclass[reprint,aps,amsmath,amssymb,superscriptaddress]{revtex4-1}
\usepackage{graphicx}
\usepackage{dcolumn}
\usepackage{bm}
\usepackage{hyperref}
\usepackage[usenames]{color}

\newcommand{\fuzhou}{Department of Physics, Fuzhou University, Fuzhou 350108, Fujian, China }
\newcommand{\fujian}{Fujian Science and Technology Innovation Laboratory for Optoelectronic Information of China, Fuzhou 350108, Fujian, China}

\begin{document}


\title{Negative differential thermal resistance of fluids induced by heat baths}



\author{Rongxiang Luo}
\email[]{phyluorx@fzu.edu.cn}
\affiliation{\fuzhou}
\affiliation{\fujian}
\author{Juncheng Guo}
\affiliation{\fuzhou}
\author{Jun Zhang}
\affiliation{\fuzhou}
\author{Hanxin Yang}
\affiliation{\fuzhou}


\date{\today}

\date{\today}
\begin{abstract}
It has recently been shown that in one-dimensional hard-point gases, there is a mechanism that induces negative differential thermal resistance (NDTR) between heat baths. We examine this mechanism in more general higher dimensional fluids described by multiparticle collision dynamics. We consider fluids in a finite cuboid region of three-dimensional space with each end in contact with a heat bath. Based on analytical results and numerical models, we find that the mechanism underlying NDTR also works for high-dimensional fluidic systems with weak interactions and is very robust to mixed fluids. Our results significantly advance knowledge of NDTR induced by heat bath and illuminate new directions to explore in fabricating fluid thermal transistors in micro- and nanosystems.\par
\end{abstract}


\maketitle



\textit{\textbf{Introduction.}} Negative differential thermal resistance (NDTR) is an important heat transport phenomenon~\cite{B. Li2006,L. Wang2007,L. Wang2008}. NDTR is observed in a system when the heat current counterintuitively decreases as the temperature difference between heat baths increases; it is analogous to electronic negative differential resistance. Our ultimate intention is to fabricate real thermal devices with the NDTR effect that enable us to control and manage heat current, thus leading to some novel and exciting applications ~\cite{N. B. Li2012}.\par

The study of NDTR at nanoscale is of fundamental theoretical interest in identifying the basic properties of heat transport for solid systems. NDTR was originally observed in nonlinear phononic lattices in 2004 by Li and colleagues~\cite{B. Li2004}. It was then exploited in the design of various thermal functional devices such as thermal transistor~\cite{B. Li2006}, thermal logic gates~\cite{L. Wang2007} and thermal memory~\cite{L. Wang2008}, among others~\cite{J. Wu2012}. NDTR has since been observed in various low-dimensional lattice models and its properties have been found to depend on the various parameters of different systems~\cite{N. Yang2007,E. Pereira2010,D. He2009,W. R. Zhong2011,Shao2009,D. Segal2006,W. Chung Lo2008,H.K. Chan2014}. At present, the NDTR phenomenon in a lattice has been understood in terms of phonon-phonon interactions and nonlinear dynamic localization of phonon modes~\cite{B. Li2006,W. C. Fu2015}. The necessary conditions for the occurrence of NDTR have already been analytically identified~\cite{H.K. Chan2014,M. S. Mendonca2015}. These significant progresses have created new knowledge for an important class of lattice systems. Besides, inspired by the above pioneer works, NDTR has also been extensively investigated in various quantum systems in order to design quantum thermal devices, typically quantum thermal transistors~\cite{P. Ben2014,K. Joulain2016,J. Ren2013,A. Fornieri2016,B. Q. Guo2018,H. Liu2019}. And also, the study along this line has turned out very successful and fruitful. \par

Fluid flow regulation at micro- and nanoscales is essential in integrated fluidic devices, which have widespread application in biology, medicine, chemistry, and engineering~\cite{T.Thorsen2002,P. S. Dittrich2006,D. Mark2010,A. Groisman2004}. Various fluidic thermal control devices (TCDs), such as fluidic thermal switches, thermal diodes and thermal regulators, have been incorporated into applications at different size scales and temperature ranges; see Ref.~\cite{K. Klinar2021} for a review and references. However, despite these achievements, an effective fluidic thermal transistor has yet to be developed. The principal reason for this situation, apart from the engineering and material challenges, stems from a long-term lack of understanding of NDTR for the fluid systems. \par

An original step towards such understanding was made in 2019 by Luo, who showed that in one-dimensional (1D) hard-point gas systems, representing 1D fluids, NDTR can be induced by heat baths at different temperatures~\cite{Luo2019}. NDTR in such a system depends on the motion of particles being weakened by decreasing the temperature of the cold bath so that collisions between the colder particles and the hot bath become very infrequent. As a result, there is little thermal current even when the temperature difference between the heat baths is large. This observed result provides a new perspective on the NDTR phenomenon and can inform the design of a fluidic thermal transistor and other more complex fluidic TCDs. However, although the mechanism is general, only a basic model (a 1D chain of hard-point elastically colliding particles) has been investigated. In order to obtain an indepth and comprehensive understanding of NDTR, the relevant questions now are: How general is this mechanism? Will this mechanism work for more general higher dimensional fluids, and how robust is it? We need the answers to these questions in order to promote the wider practical application of TCDs based on NDTR.\par

In this paper, we provide a positive answer to both the preceding questions. It might be initially convenient to consider a 3D system of fluids described by multiparticle collision dynamics (MPC)~\cite{A. Malevanets1999}. An important feature of MPC is that the velocities of conventional deterministic molecular dynamics are replaced by a set of stochastically determined velocities which satisfy the general properties of the hydrodynamic equations in numerical modeling~\cite{J.T.Padding2006}. Using this technique, researchers gained a considerable understanding of various aspects of particle transport~\cite{G. Gompper2009}. This technique has more recently been used to study the coupled particle and heat transport~\cite{Benenti2014,R.X. Luo2018} and has been instrumental in testing theoretical conjectures of heat conduction in momentum-conserving systems~\cite{P. D. Cintio2017,R. X. Luo2020,R. X. Luo2021}. In this paper, by analyzing the more general 3D MPC fluids, we show that the mechanism for NDTR induced by heat baths is also in effect for higher dimensional fluids and is suitable for describing systems with weak interactions. The universality of the mechanism is further supported by confirmation of its robustness in mixed fluids.\par

\begin{figure}
\includegraphics[width=7cm]{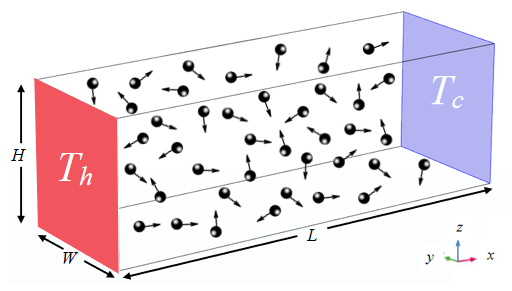}
\caption{(Color online) Schematic illustration of the 3D fluid of interacting particles in a cuboid volume described by the multi-particle collision dynamics. The system is coupled at its left and right ends to one of two heat baths at fixed temperature $T_h$ and $T_c$ (See text for more details).}
\label{fig1}
\end{figure}

\textit{\textbf{The 3D fluid model.}}  The 3D fluid model we use in this study is shown in Fig.~\ref{fig1}. The system consists of $N$ interacting point particles with equal mass $m$, with all particles are confined in a cuboid volume of length $L$, width $W$, and height $H$ in the $x$, $y$ and $z$ coordinate system shown. At $x=0$ and $x=L$ in longitudinal direction, the particles exchange heat with a heat bath of temperature $T_h$ or $T_c$; the heat baths are modeled as thermal walls ~\cite{Lebowitz1978}. When a particle arrives the $x=0$ (or $x=L$) boundary (of area $WH$), it is reflected back with a newly assigned velocity ($v_x$, $v_y$ and $v_z$ in the $x$, $y$ and $z$ directions) determined by sampling from a given distribution~\cite{Lebowitz1978}:
\begin{equation}\label{Eq1}
\begin{aligned}
  f\left(v_{x}\right)= & \frac{ m|v_{x}|}{k_{B}T_\alpha}\textrm{exp}\left(-\frac{mv^{2}_{x}}{2k_{B}T_\alpha}\right), \\
 f(v_{y,z})= &\sqrt{\frac{m}{2\pi k_{B}T_\alpha}}\textrm{exp}\left(-\frac{mv^{2}_{y,z}}{2k_{B}T_\alpha}\right),
\end{aligned}
\end{equation}
where $T_\alpha$ ($\alpha=h, c$) is the temperature of the heat bath in dimensionless units and $k_{B}$ is the Boltzmann constant. The particles are subject to periodic boundary conditions in the $y$ and $z$ directions. We point out that  the numerical results also apply to fixed boundary conditions since, in both cases, $v_{y,z}>0$ and $v_{y,z}<0$ with equal probability $p=0.5$.\par

The dynamics of the system are described by MPC~\cite{A. Malevanets1999,J.T.Padding2006,G. Gompper2009}, which simplifies the numerical modeling of particle interactions by coarse-graining the time and space at which interactions occur. In MPC, the system changes in discrete time steps, each step consisting of noninteractive propagation during a time interval $\tau$ followed by an instantaneous collision event. During propagation, the velocity  $\mathbf{v}_i$ of a particle is unchanged, and its position is updated as
\begin{equation}\label{Eq2}
\mathbf{r}_i\rightarrow\mathbf{r}_i+\tau\mathbf{v}_i.
\end{equation}
To model collisions, the system volume is partitioned into cubic cells of side $a$ and, for all particles in a cell, their velocities are rotated around a randomly chosen axis, with respect to their center of mass velocity $\mathbf{V}_{\textrm{c.m.}}$ by an angle, $\theta$ or $-\theta$, randomly chosen with equal probability $p=0.5$. The velocity of a particle in a cell is thus updated as
\begin{equation}\label{Eq3}
\mathbf{v}_i\rightarrow\mathbf{V}_{\textrm{c.m.}}+\hat{\mathcal{R}}^{\pm\theta}\left(\mathbf{v}_i-\mathbf{V}_{\textrm{c.m.}}\right),
\end{equation}
where $\hat{\mathcal{R}}^{\pm\theta}$ is the rotation operator through the angle $\pm\theta$. The movements described maintain the total momentum and energy of the fluid system. Note that the angle $\theta=\pi/2$ corresponds to the most efficient mixing of the particle momenta. Note also that the probability of collision between particles increases as $\tau$ decreases, and thus the time interval $\tau$ between successive collisions can be used to tune the strength of the interactions and consequently affect the transport of the particles. \par

In our modeling, we set $T_h=1$ and  $T_c=1-\Delta T$, where $\Delta T$ is the temperature differential of the system. Then the main parameters are set as follows: $m=k_B=W=H=1$, $a=0.1$, $\theta=\pi/2$, and the averaged particle number density $\rho=N/(LWH)=88$. To guarantee Galilean invariance of the stochastic rotation dynamics, the collision grid is shifted randomly before each collision step~\cite{T. Ihle2001}. Numerically, after the system reaches the steady state, we compute the thermal current $J$ that crosses the system according to its definition (i.e., the average energy exchanged in the unit time and unit area between particles and heat bath). The distributions of temperature $T(x)$ and particle density $\rho(x)$, where $x$ is the space variable, are similarly measured, as described in~\cite{Luo2019}. For all data points shown in the figures in this paper, the errors are $\leq 1\%$; as the error bars are smaller than the symbols, they are omitted.\par

\begin{figure}
\includegraphics[width=9cm]{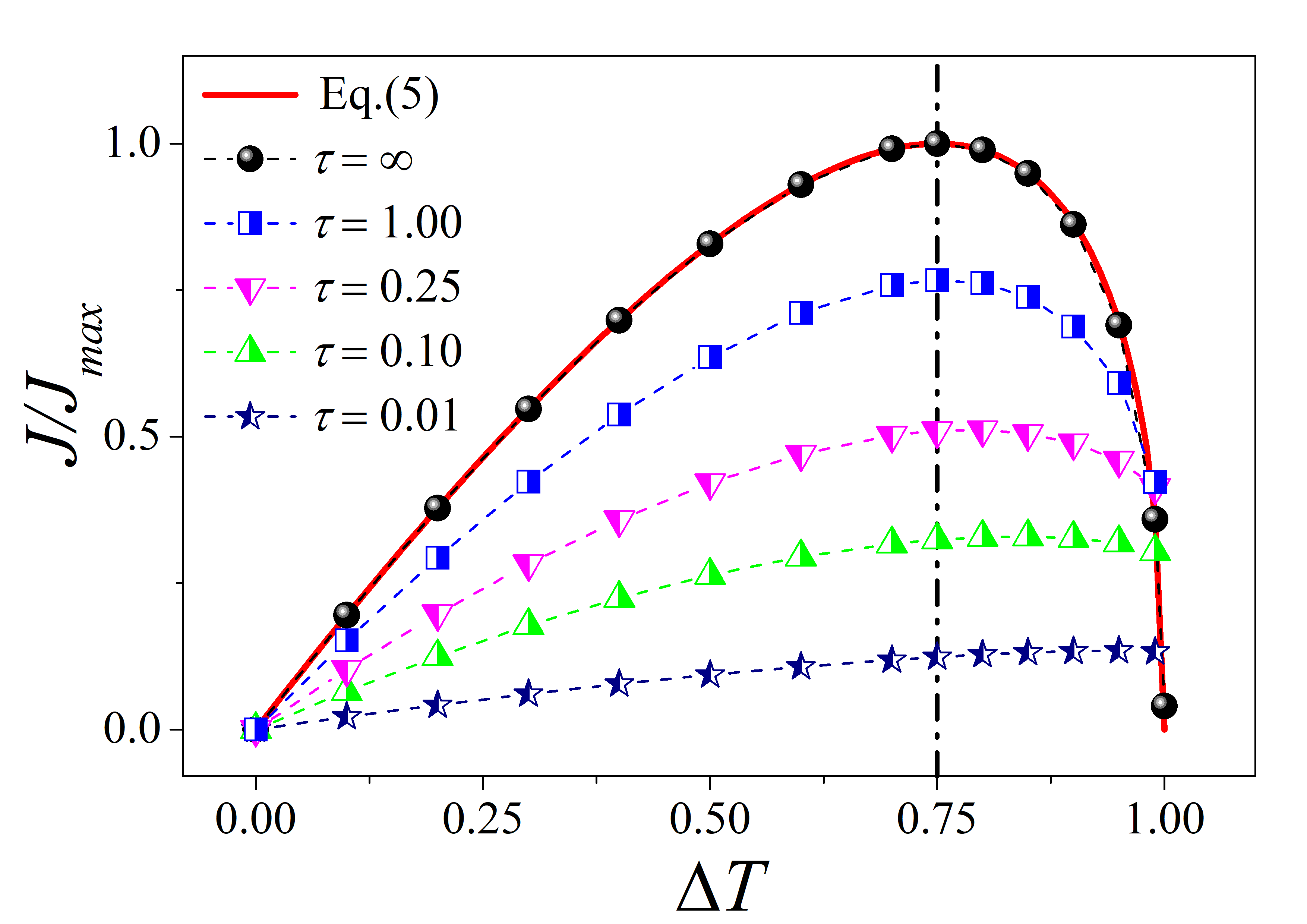}
\caption{(Color online) The thermal current $J/J_{max}$ as a function of the temperature difference $\Delta T$ for the 3D fluid system with different $\tau$ values. Here and in Fig.~\ref{fig3}, the data are obtained for $L=16$. The red line is the analytical result given by Eq.~(\ref{Eq5}). Hereafter, the symbols are modeling results and the black dot-dashed line at critical temperature difference $(\Delta T)_{cr}=0.75$ is drawn for reference. }
\label{fig2}
\end{figure}

\textit{\textbf{Analytical results.}} Here, we show that NDTR can be induced in a fluidic system by heat baths and we infer its mechanism. Note that our model is interacting and nonintegrable. However, if no particles interact (i.e., each particle maintains unchanged velocity as it crosses the system from one heat bath to the other), the model becomes integrable. In the integrable case, by an analysis similar to that performed in~\cite{Luo2019}, we obtain an analytical expression for the thermal current:

\begin{equation}\label{Eq4}
J=\left(d+1\right)\sqrt{\frac{\rho^{2}k^{3}_B}{2\pi m }}\sqrt{T_hT_c}\left(\sqrt{T_h}-\sqrt{T_c}\right),
\end{equation}
where $d$ is the spatial dimension (the analytical results presented here apply also to one- and two-dimensional systems.). To illustrate, putting $T_h=1$ and $T_c=1-\Delta T$  into Eq.~(\ref{Eq4}), we rewrite the thermal current as:
\begin{equation}\label{Eq5}
J=4J_{max}\sqrt{1-\Delta T}\left(1-\sqrt{1-\Delta T}\right).
\end{equation}
Here $J_{max}=\frac{\left(d+1\right)}{4}\sqrt{\frac{\rho^{2}k^{3}_B}{2\pi m }}$ is obtained at critical temperature difference $(\Delta T)_{cr}$, which is determined by solving the equation $\partial J/ \partial (\Delta T)=0$. This result is shown in Fig.~\ref{fig2} with a red line. It is clear that for $\Delta T>(\Delta T)_{cr}=0.75$, $J$ decreases when $\Delta T$ increases and thus exhibits the NDTR phenomenon.\par

\begin{figure}
\includegraphics[width=9cm]{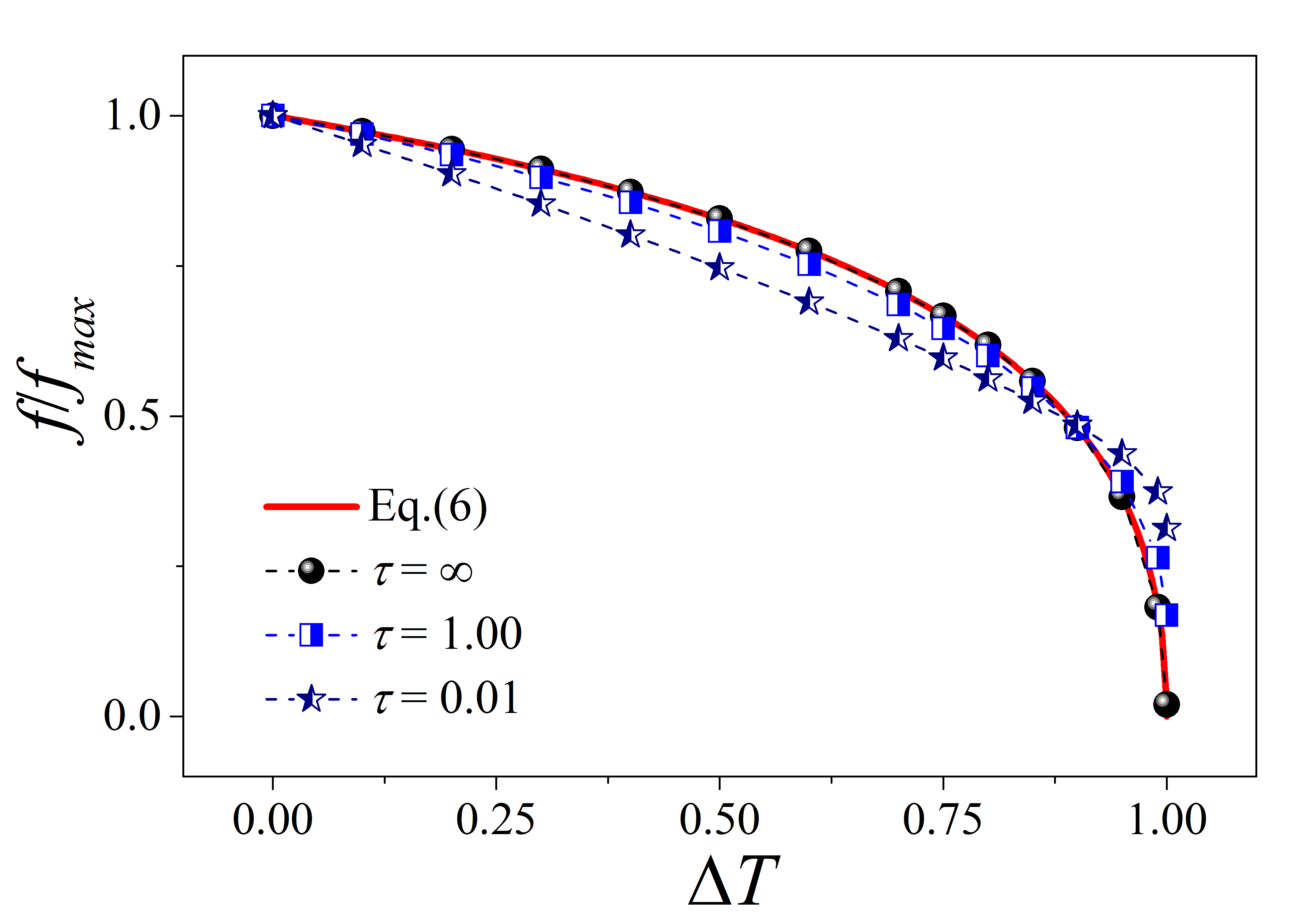}
\caption{(Color online) The collision rate $f/f_{max}$ at which the particles collide with the hot bath versus the temperature difference $\Delta T$ for the noninteracting case ($\tau=\infty$) and the interacting cases ($\tau=1.00,0.01$). Here, the red line is drawn from Eq.~(\ref{Eq6}) and $f_{max}=\frac{2N}{L}\sqrt{\frac{2k_B}{\pi m}}$.}
\label{fig3}
\end{figure}

The mechanism inferred for NDTR induced in this system can be understood from the following argument. As $T_c$ decreases (or as $\Delta T$ is increased), the particles will be reflected from the cold bath boundary at a reduced velocity; the propagation time taken by the reflected particles to return to the hot bath will increase, which in turn decreases the collision rate $f$ of particles colliding with the hot bath. To illustrate this, we can equivalently rewrite $f$, given by Eq.~(6) in Ref.~\cite{Luo2019}, as a function of $\Delta T$:

\begin{equation}\label{Eq6}
f=\frac{N}{L}\sqrt{\frac{2k_B}{\pi m}}/\left(1+\frac{1}{\sqrt{1-\Delta T}}\right).
\end{equation}
This analytical expression is also plotted with a red line in Fig.~\ref{fig3} that shows that $f$ decreases as $\Delta T$ increases, as expected. This decrease implies that $f$ will become too small for a thermal exchange between heat baths. Thus, by decreasing $T_c$ to increase $\Delta T$, the thermal exchange will be promoted in the conventional way through increasing $\Delta T$ but it will be inhibited in a new way by decreasing $f$. Both actions contribute to the thermal exchange between the baths but they compete with each other: at first, the conventional method dominates, so $J$ increases as $\Delta T$ increases; however, when $\Delta T>(\Delta T)_{cr}$, the effect of $f$ becomes dominant, so $J$ decreases as $\Delta T$ increases, thus causing the NDTR effect. The preceding analysis leads us to conclude that for the MPC fluid system, NDTR can be induced by decreasing the temperature of the heat bath.\par

\textit{\textbf{Numerical results.}} To check the analytical results and provide a numerical example, we first quantify the noninteracting system (i.e., the integrable case). In Fig.~\ref{fig2} and Fig.~\ref{fig3}, the thermal current (Eq.~(\ref{Eq5})) and the collision rate (Eq.~(\ref{Eq6})) are compared with our models (black circles). It can be seen that they agree very well with each other. These models clearly strongly support our analysis.

We now turn to the interacting systems with collisions to investigate the dependence of the mechanism on interaction strength. Here, the time interval $\tau$ between successive collisions will be used to tune the strength of the interactions. For the noninteracting case, $\tau= \infty$; thus for the interacting case, a lower value of $\tau$ produces greater interaction strength. We can see in Fig.~\ref{fig2} that although for a given system dimension ($L=16$), decreasing $\tau$ decreases the thermal current and the region of NDTR decreases in size and finally disappears, NDTR still exists for a wide range of $\tau>0.1$. This observation implies that the mechanism is more effective under relatively weak interactions. It is worth pointing out that the mechanism of NDTR for the interacting case is the same as for the noninteracting case because, as shown in Fig.~\ref{fig3}, for the interacting case ($\tau=1$), $f$ decreases as $\Delta T$ increases, as in the noninteracting case.\par

\begin{figure}
\includegraphics[width=9cm]{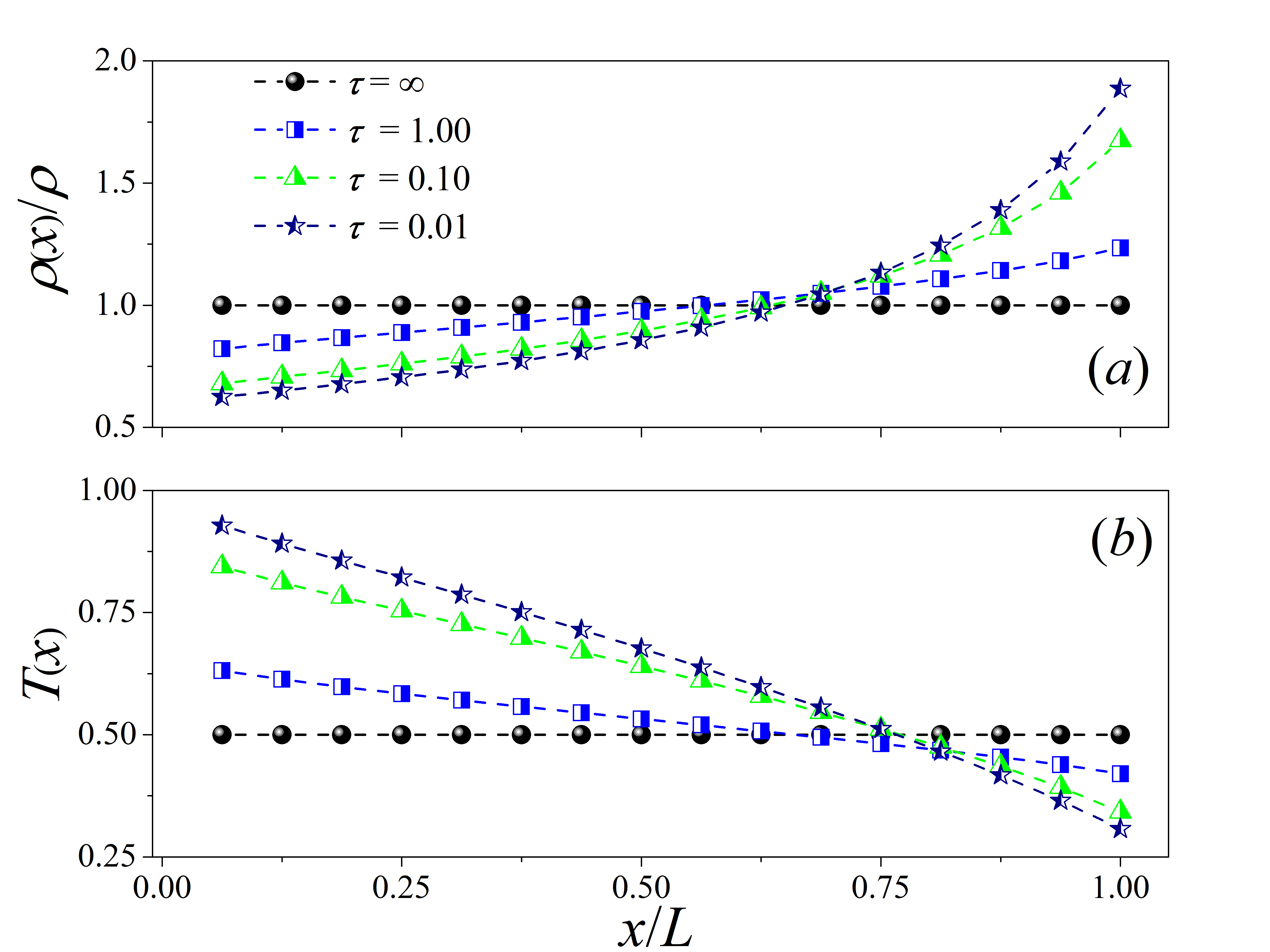}
\caption{(Color online) (a) The particle density $\rho(x)/\rho$ and (b) temperature $T(x)$ as a function of $x/L$ for different values of $\tau$. Here, we set $L=16$ and $\Delta T=0.75$. }
\label{fig4}
\end{figure}

Next, we explain why the mechanism does not work for systems with strong interactions. To this end, we plot the particle density $\rho(x)$ and temperature profile $T(x)$ of the system with different $\tau$ values. It can be seen in Fig.~\ref{fig4}(a) that $\rho(x)$ at the right (cold bath) side of the system increases as $\tau$ decreases. This implies that when the interaction strength increases, a particle with low velocity that is reflected from the cold bath will become less and less likely to transit from the cold bath to the hot bath without interacting with other particles. As a result, momentum exchange between particles will increase and low velocity particles will increase in velocity, resulting in an increase in $f$ for $\tau=0.01$ over a small range of approximately $\Delta T>0.9$, as shown in Fig.~\ref{fig3}. It also can be seen in Fig.~\ref{fig4}(b) that for lower values of $\tau$, $T(x)$ will be close to a linear response described by Fourier's law, which indicates that for $\tau=0.01$, the contribution of increasing $\Delta T$ to the thermal current again becomes dominant (see the data in Fig.~\ref{fig2} for reference). This analysis supports the inference that the mechanism will fail in the case of strong interactions.\par

\begin{figure}
\includegraphics[width=9cm]{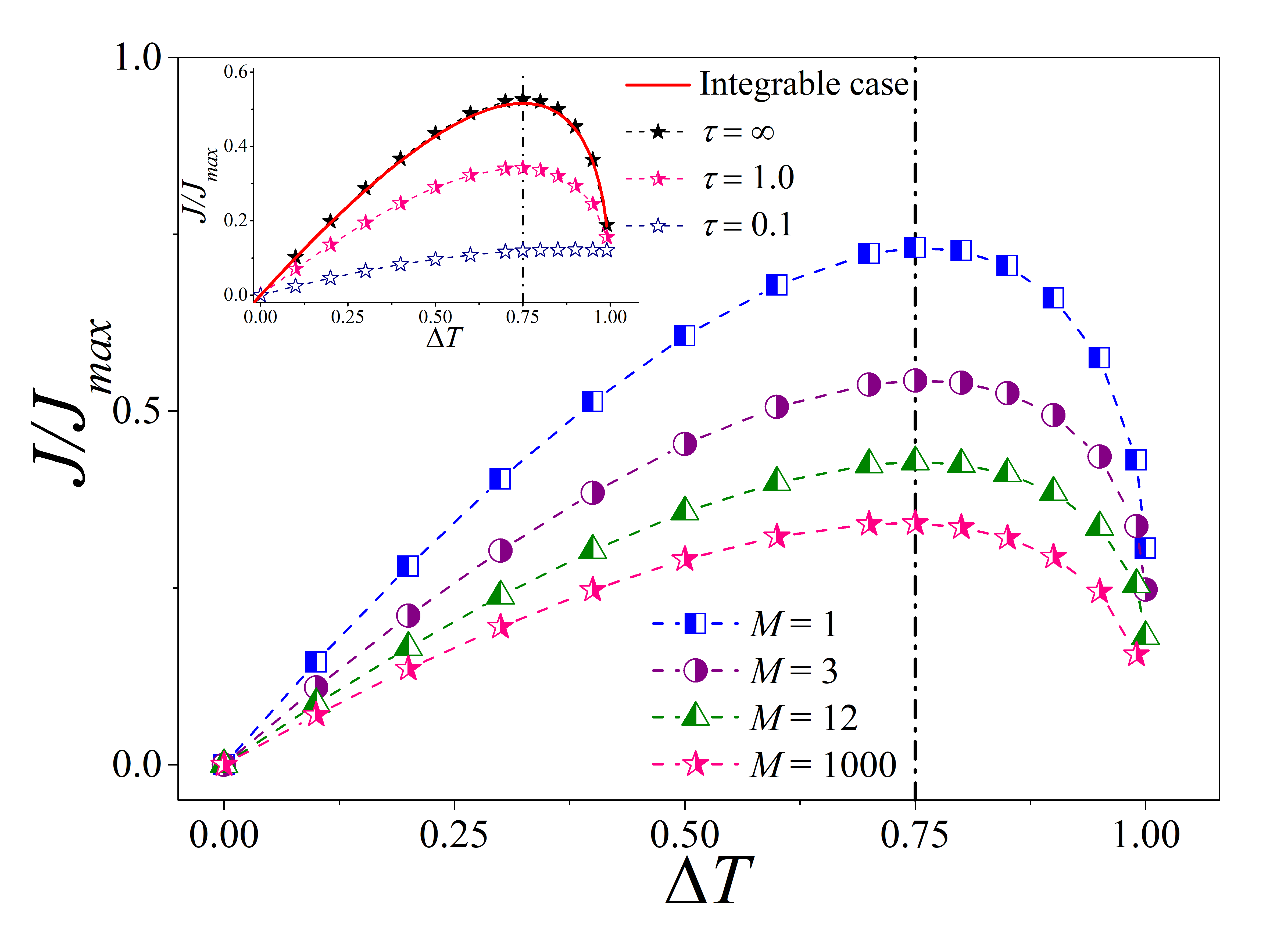}
\caption{(Color online) The thermal current $J/J_{max}$ as a function of the temperature difference $\Delta T$ for the 3D  binary fluid system with different mass ratios $M$. Here, we set $L=20$, $\tau=1.0$ and $p=0.5$. Inset: $J/J_{max}$ as a function of $\Delta T$ for $M=1000$ with different $\tau$ values. The red line in the inset is the analytical result for the integrable case ($\tau=\infty$) obtained by generalizing Eq.~(\ref{Eq4}).}
\label{fig5}
\end{figure}

To further support the universality of this mechanism, we show that it is also applicable to mixed fluids. MPC dynamics make it convenient to numerically analyze the behavior of mixed fluids by modeling it using the method described in Ref.~\cite{G. Gompper2009}. As an illustration, we consider the simple binary fluids as follows. The particles are set to two different masses, $m$ and $M$, with equal probability $p=0.5$.  For convenience, we set $m=1$ so that $M$ is the mass ratio $M/m$ of the different binary fluids. The model results for different values of $M$ are shown in Fig.~\ref{fig5}. It can be seen that the mechanism works both in a pure fluid ($M=1$) and in different mixed binary fluids. More importantly, our data show that for a given interaction strength ($\tau=1$), the NDTR region remains unchanged when $M$  is increased from 1 to 1000, which implies that this mechanism is robust for many binary fluids. In the inset of Fig.~\ref{fig5}, we emphasize that for $M=1000$, the interaction when $\tau=1$ is not too weak because the thermal current for $\tau=1$, when compared to the integrable case ($\tau=\infty$), obviously changes; once again, we see that the mechanism will break down if $\tau$ decreases further, as shown in Fig.~\ref{fig2}. In addition, we have numerically checked that for various binary fluids, using different values of $p$, the mechanism works for systems with weak interactions. These results may be helpful for understanding and controlling thermal transport of mixed fluids under specific conditions~\cite{T. M. Squires2005,R. B. Schoch2008,L. Li2015}.\par

\textit{\textbf{Summary and discussion.}} In studying a 3D fluid described by MPC dynamics, we have shown, for the first time, that NDTR can be induced by heat baths in classical fluids with weak interactions. The induced mechanism depends on the simple fact that decreasing the temperature of the cold bath weakens the motion of particles and decreases the collision rate between particles and the hot bath, thus impeding thermal exchange between cold and hot baths. We demonstrated the universality of the mechanism by showing that it is operable and very robust for various mixed fluids. The results we obtained significantly advance knowledge of NDTR induced by heat baths and clearly answer the two questions raised in the introduction.\par

We note that in nonequilibrium two-qubit systems, negative differential thermal conductance can also be induced by the strong system-bath coupling~\cite{H. Liu2019}. This observation, together with our results, provides strong evidence that heat baths are important in inducing NDTR in both quantum systems and classical fluid systems. To the best of our knowledge, there are no reports of NDTR being used to design fluidic TCDs, although a theoretical thermal transistor using gas-liquid transition has been proposed~\cite{N.Ito2011}. Thus we expect that the mechanism for NDTR we present here will be exploited to design a fluidic thermal transistor and other more complex fluidic TCDs, as was done for the lattice system~\cite{B. Li2006,L. Wang2007,L. Wang2008}. As well as being of fundamental theoretical interest, our results may find application in the context of ultracold atoms, where a thermoelectric heat engine for weakly interacting particles has already been demonstrated~\cite{J.P. Brantut2013}. Finally, we point out that the MPC fluids we used in this study is a very popular model in mesoscopic physics~\cite{G. Gompper2009} and we therefore conjecture that the mechanism can be experimentally verified in mesoscopic fluid systems. We foresee a range of interesting applications based on this work.\par

\textit{\textbf{Acknowledgments.}} We acknowledge support by the National Natural Science Foundation of China (Grant No. 12105049) and the Education Department of Fujian Province (Grant No. JAT200035).\par


\begin{thebibliography}{10}
\bibitem{B. Li2006}
B. Li, L. Wang, and G. Casati,
Negative differential thermal resistance and thermal transistor,
\href{https://aip.scitation.org/doi/abs/10.1063/1.2191730
}{\color{blue}{Appl. Phys. Lett. \textbf{88}, 143501 (2006)}}.

\bibitem{L. Wang2007}
L. Wang and B. Li,
Thermal Logic Gates: Computation with Phonons,
\href{https://journals.aps.org/prl/abstract/10.1103/PhysRevLett.99.177208}
{\color{blue}{Phys. Rev. Lett. \textbf{99}, 177208 (2007)}}.

\bibitem{L. Wang2008}
L. Wang and B. Li,
Thermal Memory: A Storage of Phononic Information,
\href{https://journals.aps.org/prl/abstract/10.1103/PhysRevLett.101.267203}
{\color{blue}{Phys. Rev. Lett. \textbf{101}, 267203 (2008)}}.

\bibitem{N. B. Li2012}
N. Li, J. Ren, L. Wang, G. Zhang, P. Hanggi, and B. Li,
Colloquium: Phononics: Manipulating heat flow with electronic analogs and beyond,
\href{https://journals.aps.org/rmp/abstract/10.1103/RevModPhys.84.1045}
{\color{blue}{Rev. Mod. Phys. \textbf{84}, 1045 (2012)}}.

\bibitem{B. Li2004}
B. Li, L. Wang, and G. Casati,
Thermal Diode: Rectification of Heat Flux,
\href{https://journals.aps.org/prl/abstract/10.1103/PhysRevLett.93.184301}
{\color{blue}{Phys. Rev. Lett. \textbf{93}, 184301 (2004)}}.

\bibitem{J. Wu2012}
J. Wu, L. Wang and B. Li,
Heat current limiter and constant heat current source,
\href{https://journals.aps.org/pre/abstract/10.1103/PhysRevE.85.061112}
{\color{blue}{Phys. Rev. E \textbf{85}, 061112 (2012)}}.

\bibitem{N. Yang2007}
N. Yang, N. Li, L. Wang, and B. Li,
Thermal rectification and negative differential thermal resistance in lattices with mass gradient,
\href{https://journals.aps.org/prb/abstract/10.1103/PhysRevB.76.020301}
{\color{blue}{Phys. Rev. B \textbf{76}, 020301(R) (2007)}}.

\bibitem{E. Pereira2010}
E. Pereira,
Graded anharmonic crystals as genuine thermal diodes: Analytical description of rectification and negative differential thermal resistance,
\href{https://journals.aps.org/pre/abstract/10.1103/PhysRevE.82.040101}
{\color{blue}{Phys. Rev. E \textbf{82}, 040101(R) (2010)}}.

\bibitem{D. He2009}
D. He, S. Buyukdagli, and B. Hu,
Origin of negative differential thermal resistance in a chain of two weakly coupled nonlinear lattices,
\href{https://journals.aps.org/prb/abstract/10.1103/PhysRevB.80.104302}
{\color{blue}{Phys. Rev. B \textbf{80}, 104302 (2009)}}.

\bibitem{W. R. Zhong2011}
W. Zhong, M. Zhang, B. Ai, and B. Hu,
Anomalous negative differential thermal resistance in a momentum-conserving lattice,
\href{https://journals.aps.org/pre/abstract/10.1103/PhysRevE.84.031130}
{\color{blue}{Phys. Rev. E \textbf{84}, 031130 (2011)}}.

\bibitem{Shao2009}
Z. Shao, L. Yang, H. Chan, and B. Hu,
Transition from the exhibition to the nonexhibition of negative differential thermal resistance in the two-segment Frenkel-Kontorova model
\href{https://journals.aps.org/pre/abstract/10.1103/PhysRevE.79.061119}
{\color{blue}{Phys. Rev. E \textbf{79}, 061119 (2009)}}.

\bibitem{D. Segal2006}
D. Segal,
Heat flow in nonlinear molecular junctions: Master equation analysis,
\href{https://journals.aps.org/prb/abstract/10.1103/PhysRevB.73.205415}
{\color{blue}{Phys. Rev. B \textbf{73}, 205415 (2006)}}.

\bibitem{W. Chung Lo2008}
W. Lo, L. Wang, and B. Li,
Thermal Transistor: Heat Flux Switching and Modulating,
\href{https://journals.jps.jp/doi/abs/10.1143/JPSJ.77.054402}
{\color{blue}{J. Phys. Soc. Jpn. \textbf{77}, 054402 (2008)}}.

\bibitem{H.K. Chan2014}
H. Chan, D. He, and B. Hu,
Scaling analysis of negative differential thermal resistance,
\href{https://journals.aps.org/pre/abstract/10.1103/PhysRevE.89.052126}
{\color{blue}{Phys. Rev. E  \textbf{89}, 052126 (2014)}}.

\bibitem{W. C. Fu2015}
W. Fu, T. Jin, D. He, and S. Qu,
Effect of dynamical localization on negative differential thermal resistance,
\href{https://www.sciencedirect.com/science/article/pii/S0378437115003076?via}
{\color{blue}{Physica A  \textbf{433}, 211 (2015)}}.

\bibitem{M. S. Mendonca2015}
M. S. Mendonca and E. Pereira,
Effective approach for anharmonic chains of oscillators: Analytical description of negative differential thermal resistance,
\href{https://www.sciencedirect.com/science/article/pii/S0375960115005344?via}
{\color{blue}{Phys. Lett. A  \textbf{379}, 1983 (2015)}}.


\bibitem{P. Ben2014}
P. Ben-Abdallah and S. A. Biehs,
Near-Field Thermal Transistor,
\href{https://journals.aps.org/prl/abstract/10.1103/PhysRevLett.112.044301}
{\color{blue}{Phys. Rev. Lett. \textbf{112}, 044301 (2014)}}.

\bibitem{K. Joulain2016}
K. Joulain, J. Drevillon, Y. Ezzahri, and J. O. Miranda,
Quantum Thermal Transistor,
\href{https://journals.aps.org/prl/abstract/10.1103/PhysRevLett.116.200601}
{\color{blue}{Phys. Rev. Lett. \textbf{116}, 200601 (2016)}}.

\bibitem{J. Ren2013}
J. Ren and J. Zhu,
Anomalous energy transport across topological insulator superconductor junctions,
\href{https://journals.aps.org/prb/abstract/10.1103/PhysRevB.87.165121}
{\color{blue}{Phys. Rev. B \textbf{87}, 165121 (2013)}}.

\bibitem{A. Fornieri2016}
A. Fornieri, G. Timossi, R. Bosisio, P. Solinas, and F. Giazotto,
Negative differential thermal conductance and heat amplification in superconducting hybrid devices,
\href{https://journals.aps.org/prb/abstract/10.1103/PhysRevB.93.134508}
{\color{blue}{Phys. Rev. B \textbf{93}, 134508 (2016)}}.

\bibitem{B. Q. Guo2018}
B. Guo, T. Liu, and C. Yu,
Quantum thermal transistor based on qubit-qutrit coupling,
\href{https://journals.aps.org/pre/abstract/10.1103/PhysRevE.98.022118}
{\color{blue}{Phys. Rev. E \textbf{98}, 022118 (2018)}}.

\bibitem{H. Liu2019}
H. Liu, C. Wang, L. Wang, and J. Ren,
Strong system-bath coupling induces negative differential thermal conductance and heat amplification in nonequilibrium two-qubit systems,
\href{https://journals.aps.org/pre/abstract/10.1103/PhysRevE.99.032114}
{\color{blue}{Phys. Rev. E \textbf{99}, 032114 (2019)}}.

\bibitem{T.Thorsen2002}
T. Thorsen, S. J. Maerkl, and S. R. Quake,
Microfluidic Large-Scale Integration,
\href{https://www.science.org/doi/10.1126/science.1076996}
{\color{blue}{Science \textbf{298}, 580 (2002)}}.

\bibitem{P. S. Dittrich2006}
P. S. Dittrich and A. Manz,
Lab-on-a-chip: microfluidics in drug discovery,
\href{https://www.nature.com/articles/nrd1985}
{\color{blue}{Nat. Rev. Drug Discov. \textbf{5}, 210 (2006)}}.

\bibitem{D. Mark2010}
D. Mark, S. Haeberle, G. Roth, F. von Stetten, and R. Zengerle,
Microfluidic lab-on-a-chip platforms: requirements, characteristics and applications,
\href{https://pubs.rsc.org/en/content/articlelanding/2010/CS/b820557b}
{\color{blue}{Chem. Soc. Rev.  \textbf{39}, 1153 (2010)}}.

\bibitem{A. Groisman2004}
A. Groisman and S. R. Quake,
A Microfluidic Rectifier: Anisotropic Flow Resistance at Low Reynolds Numbers,
\href{https://journals.aps.org/prl/abstract/10.1103/PhysRevLett.92.094501}
{\color{blue}{Phys. Rev. Lett. \textbf{92}, 094501 (2004)}}.

\bibitem{K. Klinar2021}
K. Klinar, T. Swoboda, M. M. Rojo, A. Kitanovski,
Fluidic and Mechanical Thermal Control Devices,
\href{https://onlinelibrary.wiley.com/doi/full/10.1002/aelm.202000623}
{\color{blue}{Adv. Electron. Mater. \textbf{7}, 2000623 (2021)}}.

\bibitem{Luo2019}
R. Luo,
Negative differential thermal resistance in one-dimensional hard-point gas models,
\href{https://journals.aps.org/pre/abstract/10.1103/PhysRevE.99.032138}
{\color{blue}{Phys. Rev. E \textbf{99}, 032138 (2019)}}.

\bibitem{A. Malevanets1999}
A. Malevanets and R. Kapral,
Mesoscopic model for solvent dynamics,
\href{https://aip.scitation.org/doi/abs/10.1063/1.478857}
{\color{blue}{J. Chem. Phys. \textbf{110}, 8605 (1999)}}.

\bibitem{J.T.Padding2006}
J. T. Padding and A. A. Louis,
Hydrodynamic interactions and Brownian forces in colloidal suspensions: Coarse-graining over time and length scales,
\href{https://journals.aps.org/pre/abstract/10.1103/PhysRevE.74.031402}
{\color{blue}{Phys. Rev. E \textbf{74}, 031402 (2006)}}.

\bibitem{G. Gompper2009}
G. Gompper, T. Ihle, D. M. Kroll, and R. G. Winkler,
\emph{Multi-Particle Collision Dynamics: A Particle-Based Mesoscale Simulation Approach to the Hydrodynamics of Complex Fluids}
\href{https://link.springer.com/chapter/10.1007/978-3-540-87706-6_1}
{\color{blue}{(Springer, Berlin 2009)}}.

\bibitem{Benenti2014}
G. Benenti, G. Casati, C. Mej{\'i}a-Monasterio,
Thermoelectric efficiency in momentum-conserving systems,
\href{https://iopscience.iop.org/article/10.1088/1367-2630/16/1/015014}
{\color{blue}{New J. Phys. \textbf{16}, 015014 (2014)}}.

\bibitem{R.X. Luo2018}
R. Luo, G. Benenti, G. Casati, and J. Wang,
Thermodynamic Bound on Heat-to-Power Conversion,
\href{https://journals.aps.org/prl/abstract/10.1103/PhysRevLett.121.080602}
{\color{blue}{Phys. Rev. Lett. \textbf{121}, 080602 (2018)}};
Onsager reciprocal relations with broken time-reversal symmetry,
\href{https://journals.aps.org/prresearch/abstract/10.1103/PhysRevResearch.2.022009#fulltext}
{\color{blue}{Phys. Rev. Research \textbf{2}, 022009(R) (2020)}}.

\bibitem{P. D. Cintio2017}
P. DiCintio, R. Livi, S. Lepri, and G. Ciraolo,
Multiparticle collision simulations of two-dimensional one-component plasmas: Anomalous transport and dimensional crossovers,
\href{https://journals.aps.org/pre/abstract/10.1103/PhysRevE.95.043203}
{\color{blue}{Phys. Rev. E \textbf{95}, 043203 (2017)}}.

\bibitem{R. X. Luo2020}
R. Luo,
Heat conduction in two-dimensional momentum-conserving and -nonconserving gases,
\href{https://journals.aps.org/pre/abstract/10.1103/PhysRevE.102.052104}
{\color{blue}{Phys. Rev. E \textbf{102}, 052104 (2020)}}.

\bibitem{R. X. Luo2021}
R. Luo, L. Huang, and S. Lepri,
Heat conduction in a three-dimensional momentum-conserving fluid,
\href{https://journals.aps.org/pre/abstract/10.1103/PhysRevE.103.L050102?ft=1#fulltext}
{\color{blue}{Phys. Rev. E \textbf{103}, L050102 (2021)}}.

\bibitem{Lebowitz1978}
J. L. Lebowitz and H. Spohn,
Transport properties of the Lorentz gas: Fourier's law,
\href{https://link.springer.com/article/10.1007/BF01011774}
{\color{blue}{J. Stat. Phys. \textbf{19}, 633 (1978)}};
R. Tehver, F. Toigo, J. Koplik, and J. R. Banavar,
Thermal walls in computer simulations,
\href{https://journals.aps.org/pre/abstract/10.1103/PhysRevE.57.R17}
{\color{blue}{Phys. Rev. E \textbf{ 57}, 17(R) (1998)}}.

\bibitem{T. Ihle2001}
T. Ihle and D. M. Kroll,
Stochastic rotation dynamics: A Galilean-invariant mesoscopic model for fluid flow,
\href{https://journals.aps.org/pre/abstract/10.1103/PhysRevE.63.020201}
{\color{blue}{Phys. Rev. E \textbf{63}, 020201(R) (2001)}}.

\bibitem{T. M. Squires2005}
T. M. Squires and S. R. Quake,
Microfluidics: Fluid physics at the nanoliter scale,
\href{https://journals.aps.org/rmp/abstract/10.1103/RevModPhys.77.977}
{\color{blue}{Rev. Mod. Phys. \textbf{77}, 977 (2005)}}.

\bibitem{R. B. Schoch2008}
R. B. Schoch, J. Han, and P. Renaud,
Transport phenomena in nanofluidics,
\href{https://journals.aps.org/rmp/abstract/10.1103/RevModPhys.80.839}
{\color{blue}{Rev. Mod. Phys. \textbf{80}, 839 (2008)}}.

\bibitem{L. Li2015}
L. Li, J. Mo, and Z. Li,
Nanofluidic Diode for Simple Fluids without Moving Parts,
\href{https://journals.aps.org/prl/abstract/10.1103/PhysRevLett.115.134503}
{\color{blue}{Phys. Rev. Lett. \textbf{115}, 134503 (2015)}}.

\bibitem{N.Ito2011}
T. S. Komatsu and N. Ito
Thermal transistor utilizing gas-liquid transition,
\href{https://journals.aps.org/pre/abstract/10.1103/PhysRevE.83.012104}
{\color{blue}{Phys. Rev. E \textbf{83}, 012104 (2011)}}.

\bibitem{J.P. Brantut2013}
J. P. Brantut, C. Grenier, J. Meineke, D. Stadler, S.Krinner, C. Kollath, T. Esslinger, and A. Georges,
A Thermoelectric Heat Engine with Ultracold Atoms,
\href{https://www.science.org/doi/10.1126/science.1242308}
{\color{blue}{Science \textbf{342}, 713 (2013)}}.


\end{thebibliography}
\end{document}